\documentclass[showpacs,preprintnumbers,amsmath,amssymb]{revtex4}
\usepackage{graphicx}
\usepackage{dcolumn}
\usepackage{bm}

\begin{document}

\title{EMERGENT UNIVERSE FROM SCALE INVARIANT TWO MEASURES THEORY }

\author{Sergio del Campo}
\email{sdelcamp@ucv.cl} \affiliation{ Instituto de F\'{\i}sica,
Pontificia Universidad Cat\'{o}lica de Valpara\'{\i}so, Avenida
Brasil 2950, Casilla 4059, Valpara\'{\i}so, Chile.}
\author{Eduardo I. Guendelman}
\email{guendel@bgu.ac.il} \affiliation{ Physics Department, Ben
Gurion University of the Negev, Beer Sheva 84105, Israel}
\author{Ram\'on Herrera}
\email{ramon.herrera@ucv.cl} \affiliation{ Instituto de
F\'{\i}sica, Pontificia Universidad Cat\'{o}lica de
Valpara\'{\i}so ,  Avenida Brasil 2950, Casilla 4059,
Valpara\'{\i}so, Chile.}
\author{Alexander B. Kaganovich}
\email{alexk@bgu.ac.il} \affiliation{ Physics Department, Ben
Gurion University of the Negev, Beer Sheva 84105, Israel}
\author{Pedro Labra$\tilde{n}$a}
\email{plabrana@ubiobio.cl} \affiliation{ Departamento de
F\'{\i}sica, Universidad  del B\'{\i}o B\'{\i}o, Avenida Collao
1202, Casilla 5-C, Concepci\'on, Chile.}


\date{\today}

\begin{abstract}

The dilaton-gravity sector of a linear in the scalar curvature,
scale invariant Two Measures Field Theory (TMT), is explored in
detail in the context of closed FRW cosmology and shown to allow
stable emerging universe solutions. The model possesses scale
invariance which is spontaneously broken due to the intrinsic
features of the TMT dynamics.  We study  the transition from the
emerging phase to inflation, and then to a zero cosmological
constant phase. We also study the spectrum of density
perturbations and the constraints that impose on the parameters of
the theory.
\end{abstract}

\maketitle


As a way to address the cosmological constant (CC) problem \cite{Weinberg1}-\cite{CC},
the accelerated expansion of the late time universe\cite{accel},
the cosmic coincidence \cite{coinc} (see also reviews on dark
energy\cite{de-review}-\cite{Copeland}, dark matter
\cite{dm-review} and references therein), many models have
been proposed with the aim to find answer to these puzzles, for
example: the quintessence\cite{quint},  coupled
quintessence\cite{Amendola},
$k$-essence\cite{k-essence},\cite{Caldwell-Steinhardt-Mukhanov}.

One can add to the list of puzzles the problem of initial
singularity\cite{singular},\cite{HE-book}, including the
singularity theorems for scalar field-driven inflationary
cosmology\cite{Borde-Vilenkin-PRL}, resolution of which is perhaps
a crucial criteria for the true theory. The avoidance of the
initial singularity  will be the central question that we will
address in this paper, exploring the idea of the "emerging
universe", where the universe has a non singular origin, such that
the Einstein Universe. Although the original proposal for the
emerging Universe \cite{orgemm} suffered an instability, several
proposals to formulate a stable model have been given \cite{emm2},
in particular one is obtained by invoking Jordan Brans Dicke
models\cite{emm3}

In this paper we explore a model including gravity and a single
scalar field $\phi$ in the framework of the so called Two Measures
Field Theory (TMT)\cite{GK1}-\cite{GK8}. In TMT,  many
cosmological issues can be addressed: zero vacuum energy is
obtained without fine tuning,
 the fifth force problem is resolved and
the Einstein's GR is restored when the local fermion matter energy
density (i.e in the space-time regions occupied by matter) is much
larger than the vacuum energy density. In this paper, we will
address the existence and stability of the emerging universe in
TMT. In a previous  work, \cite{prev} in which a   square
curvature term was almost of the TMT structure (except for a
contribution that gave rise to a cosmological term), an emergent
universe was described.

TMT is a generally coordinate invariant theory, where  the  action
has to be of the form\cite{GK3}-\cite{GK8}
\begin{equation}
    S = \int L_{1}\Phi d^{4}x +\int L_{2}\sqrt{-g}d^{4}x,
\label{S}
\end{equation}
including two Lagrangians $ L_{1}$ and $L_{2}$ and two measures of
integration $\sqrt{-g}$ and $\Phi$. One is the usual measure of
integration $\sqrt{-g}$ in the 4-dimensional space-time manifold
equipped with the metric
 $g_{\mu\nu}$. The other  is the new measure of integration $\Phi$ in the same
4-dimensional space-time manifold. The measure  $\Phi$ being  a
scalar density and a total derivative (see Ref.\cite{Mstring})
 may be defined by means of  four scalar fields $\varphi_{a}$
($a=1,2,3,4$),
\begin{equation}
\Phi
=\varepsilon^{\mu\nu\alpha\beta}\varepsilon_{abcd}\partial_{\mu}\varphi_{a}
\partial_{\nu}\varphi_{b}\partial_{\alpha}\varphi_{c}
\partial_{\beta}\varphi_{d}.
\label{Phi}
\end{equation}

 It is assumed that the Lagrangian densities $ L_{1}$ and
$L_{2}$ are functions of all matter fields, the dilaton field, the
metric, the connection
 but not of the
"measure fields" ($\varphi_{a}$ ). In
such a case, i.e. when the measure fields  enter in the theory
only via the measure $\Phi$,
  the action (\ref{S}) possesses
an infinite dimensional symmetry. In the case given by
Eq.(\ref{Phi}) these symmetry transformations have the form
$\varphi_{a}\rightarrow\varphi_{a}+f_{a}(L_{1})$, where
$f_{a}(L_{1})$ are arbitrary functions of  $L_{1}$ (see details in
Ref.\cite{GK3}). In this paper we will insist in keeping this symmetry and therefore
the TMT structure.

We assume here that all fields, including also the metric,
connection and the {\it measure fields} are independent dynamical
variables. All the relations among  them are results of the
equations of motion. In particular, the independence of the metric
and the connection means that we proceed in the first order
formalism and the relation between connection and metric is not
necessarily according to Riemannian geometry.

Varying the measure fields $\varphi_{a}$, we get
$B^{\mu}_{a}\partial_{\mu}L_{1}=0 $ where
$B^{\mu}_{a}=\varepsilon^{\mu\nu\alpha\beta}\varepsilon_{abcd}
\partial_{\nu}\varphi_{b}\partial_{\alpha}\varphi_{c}
\partial_{\beta}\varphi_{d}.\label{varphiB}$
Since $Det (B^{\mu}_{a}) = \frac{4^{-4}}{4!}\Phi^{3}$ it follows
that if $\Phi\neq 0$,
\begin{equation}
 L_{1}=sM^{4} =const,
\label{varphi}
\end{equation}
where $s=\pm 1$ and $M$ is a constant of integration with the
dimension of mass.

We proceed now  to discuss the question of scale invariance in the
context of TMT. A dilaton field $\phi$ allows to realize a
spontaneously broken global scale invariance\cite{G1}. We
postulate that the theory is invariant under the global scale
transformations:
\begin{equation}
    g_{\mu\nu}\rightarrow e^{\theta }g_{\mu\nu}, \quad
\Gamma^{\mu}_{\alpha\beta}\rightarrow \Gamma^{\mu}_{\alpha\beta},
\quad \varphi_{a}\rightarrow \lambda_{ab}\varphi_{b}\quad
\text{where} \quad \det(\lambda_{ab})=e^{2\theta}, \quad
\phi\rightarrow \phi-\frac{M_{p}}{\alpha}\theta . \label{st}
\end{equation}

We choose an action which, except for the modification of the
general structure caused by the basic assumptions of TMT,
 {\it does not contain
 any exotic terms and  fields} as like as in the conventional formulation
 of the minimally coupled scalar-gravity system.
Keeping the general structure (\ref{S}), it is convenient to
represent the underlying action of our model in the following
form \cite{GKKess}:
\begin{eqnarray}
&S=&\int d^{4}x e^{\alpha\phi /M_{p}}
\left[-\frac{1}{2\,\kappa}R(\Gamma ,g)(\Phi +b_{g}\sqrt{-g})+(\Phi
+b_{\phi}\sqrt{-g})\frac{1}{2}g^{\mu\nu}\phi_{,\mu}\phi_{,\nu}-e^{\alpha\phi
/M_{p}}\left(\Phi V_{1} +\sqrt{-g}V_{2}\right)\right],
 \label{totaction}
\end{eqnarray}
where $b_g$ represents the coupling constant of the curvature
scalar and $b_\phi$ of the scalar kinetic term to $\sqrt{-g}$,
respectively.

We use $\kappa =8\pi/M_p^2 $ where $M_p$ is the four-dimensional
Planck mass. In the equations of motion following  from this
action, we change the metric to the new one
\begin{equation}
\tilde{g}_{\mu\nu}=e^{\alpha\phi/M_{p}}(\zeta +b_{g})g_{\mu\nu},
\label{ct}
\end{equation}
where $\zeta \equiv\frac{\Phi}{\sqrt{-g}} \label{zeta}$. The
conformal metric $\tilde{g}_{\mu\nu}$  represents  the "Einstein
frame", since the connection  becomes Riemannian. Notice that
$\tilde{g}_{\mu\nu}$ is invariant under the scale transformations
(\ref{st}). After the change of variables  to the Einstein frame
the gravitational equations take the standard GR form
\begin{equation}
G_{\mu\nu}(\tilde{g}_{\alpha\beta})=\kappa\,T_{\mu\nu}^{eff},
 \label{gef}
\end{equation}
where  $G_{\mu\nu}(\tilde{g}_{\alpha\beta})$ is the Einstein
tensor.  The energy-momentum tensor, $T_{\mu\nu}^{eff}$, becomes
\begin{eqnarray}
T_{\mu\nu}^{eff}&=&\frac{\zeta +b_{\phi}}{\zeta +b_{g}}
\left(\phi_{,\mu}\phi_{,\nu}-\frac{1}{2}
\tilde{g}_{\mu\nu}\tilde{g}^{\alpha\beta}\phi_{,\alpha}\phi_{,\beta}\right)
-\tilde{g}_{\mu\nu}\frac{b_{g}-b_{\phi}}{2(\zeta +b_{g})}
\tilde{g}^{\alpha\beta}\phi_{,\alpha}\phi_{,\beta}
+\tilde{g}_{\mu\nu}V_{eff}(\phi;\zeta,M),
 \label{Tmn}
\end{eqnarray}
where the function $V_{eff}(\phi;\zeta,M)$ is defined as
following:
\begin{equation}
V_{eff}(\phi;\zeta ,M)=
\frac{b_{g}\left[sM^{4}e^{-2\alpha\phi/M_{p}}+V_{1}\right]
-V_{2}}{(\zeta +b_{g})^{2}}, \label{Veff1}
\end{equation}
where $V_1$ and $V_2$ are two arbitrary constants. In order to
have $V_{eff}>0$ the two constant, $V_1$ and $V_2$, should satisfy
the inequality $b_gV_1>V_2$ for $\phi\rightarrow\infty$.

The scalar field $\zeta$  is
determined by the consistency of (\ref{gef}) with (\ref{varphi}), which lead to the constraint
\begin{eqnarray}
&&(b_{g}-\zeta)\left[sM^{4}e^{-2\alpha\phi/M_{p}}+
V_{1}\right]-2V_{2}-\delta\cdot b_{g}(\zeta +b_{g})Z
=0,\label{constraint2}
\end{eqnarray}
where
$Z\equiv\frac{1}{2}\tilde{g}^{\alpha\beta}\phi_{,\alpha}\phi_{,\beta}$
and $\delta =\frac{b_{g}-b_{\phi}}{b_{g}}$.

The effective energy-momentum tensor (\ref{Tmn}) can be
represented in a form of  a perfect fluid $T_{\mu\nu}^{eff}=(\rho
+p)u_{\mu}u_{\nu}-p\tilde{g}_{\mu\nu}$, where
$u_{\mu}=\frac{\phi_{,\mu}}{(2Z)^{1/2}}$ with the following energy
and pressure densities resulting from Eqs.(\ref{Tmn}) and
(\ref{Veff1}) after inserting the solution $\zeta
=\zeta(\phi,Z;M)$ of Eq.(\ref{constraint2})
\begin{equation}
\rho(\phi,Z;M) =Z+ \frac{(sM^{4}e^{-2\alpha\phi/M_{p}}+V_{1})^{2}-
2\delta b_{g}(sM^{4}e^{-2\alpha\phi/M_{p}}+V_{1})Z -3\delta^{2}
b_{g}^{2}Z^2}{4[b_{g}(sM^{4}e^{-2\alpha\phi/M_{p}}+V_{1})-V_{2}]},
\label{rho1}
\end{equation}
and
\begin{equation}
p(\phi,Z;M) =Z- \frac{\left(sM^{4}e^{-2\alpha\phi/M_{p}}+V_{1}+
\delta b_{g}Z\right)^2}
{4[b_{g}(sM^{4}e^{-2\alpha\phi/M_{p}}+V_{1})-V_{2}]}. \label{p1}
\end{equation}
Notice that if $s$ and $V_{1}$ have different signs one might
obtains a state with zero energy density.

We now want to consider the detailed analysis of The Emergent
universe solutions and in the next section their stability in the
TMT scale invariant theory. We start considering the
Friedmann-Robertson-Walker  closed cosmological solutions of the
form
\begin{equation}
ds^2 =dt^2 - a(t)^2 \left(\frac{dr^2}{1 -r^2}+ r^2(d\theta^2
+sin^2\theta d\phi^2)\right),   \phi = \phi(t),
\end{equation}
where $a(t)$ is the scale factor, and the scalar field $\phi$ is a
function of the cosmic time $t$ only, due to homogeneously and
isotropy.
We will consider a scenario where the scalar field $\phi$ is
moving in the extreme right region $\phi \rightarrow \infty  $. In
this case, the expressions for the energy density $\rho$ and
pressure $p$ are given by,
\begin{equation}\label{eq.density}
\rho = \frac{A}{2} \dot{\phi}^2 + 3B\dot{\phi}^4 + C,
\end{equation}
and
\begin{equation}
p = \frac{A}{2} \dot{\phi}^2 +B\dot{\phi}^4 - C,\label{presion}
\end{equation}
respectively. Here, the constants $A,B$ and $C$ are given by,

\begin{eqnarray}
A = 1- \frac{2\delta b_g V_1}{4(b_g V_1 - V_2)}\,,\,\,\,\,
B = -\frac{\delta^2 b^2_g }{4(b_g V_1 - V_2)}\,,\,\,\,\,
\mbox{and}\,\,\,\,
C = \frac{ V^2_1}{4(b_g V_1 - V_2)}\,\label{C},
\end{eqnarray}
respectively.

It is interesting to notice that all terms proportional to $\dot{\phi}^4$ behave like "radiation", since
 $p_{\dot{\phi}^4} = \frac{\rho_{\dot{\phi}^4} }{3}$ is satisfied. In the same way, the terms proportional to $\dot{\phi}^2$ behave like "stiff"
 matter, since $p_{\dot{\phi}^2}=\rho_{\dot{\phi}^2}$, and finally, the $C$ constant term behaves like a "cosmological
 constant". The emerging universe can turn into inflation only if
 $C>0$.

The equations that determines the static closed universe $a(t) =
a_0 =constant$, in which $\dot{a}=0$, $\ddot{a}=0$, gives rise to
a restriction for $\dot{\phi}_0$ that have to satisfy  in order to
guarantee that the universe be static. Since, $\ddot{a}=0$ is
proportional to $\rho + 3p$, we must require that $\rho + 3p = 0$,
which leads to

\begin{equation}\label{e1}
3B\dot{\phi}^4_0 + A\dot{\phi}^2_0 - C=0.
\end{equation}

This equation leads to two roots,  given by

\begin{equation}\label{e2}
\dot{\phi}_{1,2}^2=\frac{\pm\sqrt{A^2+ 12BC}\,-A}{6B}\,.
\end{equation}


Defining the variable $y=\frac{2\delta b_g C}{V_1}$, we see that
$A=1-y$, $BC= -\frac{y^2}{16}$, so the condition that the
discriminant be positive, i.e., that $A^2+ 12BC>0$ gives $(1-y)^2-
\frac{3y^2}{4}>0$ or $\frac{y^2}{4}-2y +1>0$, which is satisfied
for $y<2(2-\sqrt{3})=0.54$ or $y>2(2+\sqrt{3})=7.46$. However, for
$C>0$ the constraint $\dot{\phi}^2>0$ is only satisfied for
$y<0.54$ and the other region $y>7.46$ has to be discarded.
Stability provides  further constraints. In fact, as we will see,
the second solution can never be stable.

It is also interesting to see that if the discriminant is positive
the first solution yields   automatically  a positive energy
density, if we require $C>0$. The same requirement is to be
adopted if we want the emerging solution to be able to turn into
an inflationary solution. Since $C>0$, we get that
$(b_gV_1-V_2)>0$ in agrement with the fact that $V_{eff}>0$ as was
mentioned previously,  and therefore we get that $B<0$. One can
see that the condition $\rho
>0$ for the first solution reduces to the inequality $w>
(1-\sqrt{1-w} )/2$, where $w =-12BC/A^2 >0$ (recall that $B<0$),
and as long as $w<1$, it is always true that this inequality is
satisfied.

In the following we will study the stability of the static
solution. Let us consider the perturbation equations. Considering
small deviations for $\dot{\phi}$  from the static emergent
solution  ($\dot{\phi}_0$) and also  the perturbation of the scale
factor $a$, from Eq.~(\ref{eq.density}) we obtain that

\begin{equation}\label{eq.density-pert.}
\delta \rho = A \dot{\phi}_0 \delta \dot{\phi} + 12B \dot{\phi}_0^3 \delta \dot{\phi}.
\end{equation}

At the same time $\delta \rho$ can be obtained from the perturbation of the Friedmann equation, i.e.

\begin{equation}\label{Fried.eq.}
3\left(\frac{1}{a^2}+H^2\right)=\kappa \rho,
\end{equation}
and since we are perturbing a solution which is static, i.e.,
$H=0$, we obtain that
\begin{equation}\label{pert.Fried.eq.}
-\frac{6}{a_0^3}\delta a =\kappa \delta \rho.
\end{equation}

We also have the second order Friedmann equation,

\begin{equation}\label{Fried.eq.2}
\frac{1+\dot{a}^2 + 2a\ddot{a}}{a^2}=-\kappa p.
\end{equation}

Applying to this equation the static emergent solution, i.e.
$p_0=-\rho_0/3$ and  $a=a_0$,  we get
\begin{equation}
\frac{2}{a_0^2} = -2\kappa p_0 = \frac{2}{3}\kappa \rho_0= \Omega_0 \kappa \rho_0,
\end{equation}
where we have chosen to express our result in terms of $\Omega_0$, defined by $p_0=(\Omega_0-1)\rho_0$, which for the emerging
solution has the value $\Omega_0=\frac{2}{3}$. Using this in Eq.(\ref{pert.Fried.eq.}), we obtain
\begin{equation}\label{pert.Fried.eq.2}
\delta \rho = -\frac{3\Omega_0 \rho_0}{a_0}\delta a,
\end{equation}
and equating the values of $\delta \rho$ as given by Eqs.(\ref{eq.density-pert.}) and (\ref{pert.Fried.eq.2}) we obtain a linear relation between
$\delta \dot{\phi}$ and $\delta a$, which is given by
\begin{equation}\label{delta-delta}
\delta \dot{\phi}=D_0\delta a,
\end{equation}
where

\begin{equation}
D_0 = -\frac{3\Omega_0 \rho_0}{a_0 \dot{\phi}_0 (A + 12 B \dot{\phi}_0^2)}.
\end{equation}

We now consider the perturbation of the Eq.(\ref{Fried.eq.2}). In the right hand side of this equation we take
that $p=(\Omega-1)\rho$, with
\begin{equation}\label{Omega-eq.}
\Omega = 2\Big(1 - \frac{U_{eff}}{\rho}\Big),
\end{equation}
where
\begin{equation}\label{V-eq.}
U_{eff} =C + B\,\dot{\phi}^4,
\end{equation}
and thus,  the perturbation of the Eq.(\ref{pert.Fried.eq.2}) leads to,

\begin{equation}\label{pert.Fried.eq.21}
-\frac{2\delta a}{a_0^3}+2\frac{\delta\ddot{a}}{a_0}=-\kappa \delta p =-\kappa \delta [(\Omega-1)\rho].
\end{equation}

In order to evaluate this, we use Eqs.(\ref{Omega-eq.}) and
(\ref{V-eq.}), and the expressions that relate the variations in
$a$ and $\dot{\phi}$ given by Eq.(\ref{delta-delta}).  Defining
the "small"  variable $\beta$ as ($\beta\ll 1$)

\begin{equation}
a(t) = a_0( 1+ \beta),
\end{equation}
we obtain
\begin{equation}
2\ddot{\beta}(t) + W_0^2\beta(t) = 0\,,
\end{equation}
where
\begin{equation}
W_0^2 = \Omega_0\,\rho_0\left[ \frac{24\,B\,\dot{\phi}_0^2}{A +
12\,\dot{\phi}_0^2\,B }  -6\frac{(C + B\,
\dot{\phi}_0^4)}{\rho_0} -3\kappa \Omega_0 + 2\kappa \right].
\end{equation}

Notice that the sum of the last two terms in the expression for $W_0^2$, that is $-3\kappa \Omega_0 + 2\kappa $  vanish
since $\Omega_0=\frac{2}{3}$. For the same reason, we have that $6\frac{(C + B\,\dot{\phi}_0^4)}{\rho_0} = 4$, which brings us to the simplified expression
\begin{equation}
W_0^2 =4\, \Omega_0\,\rho_0\left[ \frac{6\,B\,\dot{\phi}_0^2}{A +
12\,\dot{\phi}_0^2\,B }  - 1 \right].
\end{equation}

For the stability of the static solution, we need that  $W_0^2
>0$, where $\dot{\phi}_0^2$ is defined either by Eq.~(\ref{e2})
($\dot{\phi}_0^2=\dot{\phi}_1^2$) or
($\dot{\phi}_0^2=\dot{\phi}_2^2$). Notice that since $B<0$,
$24\,B\,\dot{\phi}_0^2<0$, so that for $W_0^2>0$, we need $A +
12\,\dot{\phi}_0^2\,B<0$.
 $W_0^2>0$ implies $\frac{6\,B\,\dot{\phi}_0^2}{A + 12\,\dot{\phi}_0^2\,B }> 1 $. Multiplying this inequality by the negative
quantity $A + 12\,\dot{\phi}_0^2\,B<0$ and evaluating separately
for $\dot{\phi}_0^2=\dot{\phi}_1^2$ and
$\dot{\phi}_0^2=\dot{\phi}_2^2$ we get that the condition
$W_0^2>0$ becomes $4\sqrt{A^2+ 12BC}<8\sqrt{A^2+ 12BC}$ if we use
the first solution and $- 4\sqrt{A^2+ 12BC}<-8\sqrt{A^2+ 12BC}$ if
we use the second solution. Of course this means  that the second
solution can never be consistent with stability. For the first
solution, we still have to verify that $A +
12\,\dot{\phi}_0^2\,B<0$. Introducing the relevant expression for
$\dot{\phi}_0^2$ appropriate for the first solution, and using
again the variable $y=\frac{2\delta b_g C}{V_1}$,  we obtain now
that $y>1/2$. Putting  all together, the existence, the stability
previously defined, as well as a reasonable cosmological picture
after the emerging phase (inflation), are satisfied for $y<1$, to
avoid negative kinetic terms during the slow roll phase of
inflation. Therefore, in order to have a picture in which the
emerge universe is stable and then pass to an inflationary phase
is obtained if the following range is satisfied
\begin{equation}
0.5<y<0.54.
\end{equation}


The study of the stability of the static solution and the
properties of the different equilibrium points could be done in a
more systematic way by using a dynamical system approach.
In this scheme we rewrite the Friedmann  and the conservation of
energy equations as an autonomous system in terms of the variables
$H=\dot{a}/a$ and $x \equiv \dot{\phi}^2$.
In order to do so, we differentiate  the Fridmann equation and
after using the expressions for $\rho$ and $p$, given by
Eqs.(\ref{eq.density}),  (\ref{presion}) and (\ref{Fried.eq.2}),
we obtain:

\begin{eqnarray}
\dot{H} &=& \frac{\kappa}{3}\Big[C - 3B\, x^2 - A\,x\Big] - H^2,
\label{dinamic1}
\end{eqnarray}
and
\begin{eqnarray}
\dot{x} &=& - 6\,H\,\frac{A\,x + 4B\,x^2}{A + 12B\,x}\,\,,
\label{dinamic2}
\end{eqnarray}
where $A = 1 - y$, as before.
The equations (\ref{dinamic1}) and (\ref{dinamic2}) are a
two-dimensional autonomous system on the variables $H$ and $x$.

In order to study the stability of the static solutions we look
for critical points of the system (\ref{dinamic1}) and
(\ref{dinamic2}). These points are

\begin{eqnarray}
\Big\{ H = 0\,, \,\,x = \frac{-A+\sqrt{A^2+12 B C}}{6 B}\Big\} ;
\label{crit1}\\
\Big\{ H = 0\,,
\,\,x = \frac{-A-\sqrt{A^2+12 B C}}{6 B} \Big\}\,; \label{crit2}\\
\Big\{ H = - \sqrt{\frac{C\kappa}{3}} \,, \,\,x = 0 \Big\} ;
\Big\{ H = \sqrt{\frac{C\kappa}{3}} \,, \,\,x = 0 \Big\}\,;
\Big\{ H = -\frac{\sqrt{\frac{(A^2+16 B C)\kappa}{B}}}{4 \sqrt{3}}
\,, \,\,x = -\frac{A}{4 B}\Big\} ;
\\
\Big\{ H = \frac{\sqrt{\frac{(A^2+16 B C)\kappa}{B}}}{4 \sqrt{3}}
, x = -\frac{A}{4 B}\Big\}\,. \label{crit3}
\end{eqnarray}

The critical points have different properties depending on the
values of the parameters of the model ($B$ and $C$). At this
moment we are not going to give an exhaustive description of these
properties for all the critical points, instead, we are going to
focus on the particular critical points which are related with
static universe. From the definition of the variables $H$ and $x$
we can note that only the first two critical points
Eqs.~(\ref{crit1}) and (\ref{crit2}) correspond to a static
universe.
In order to study the nature of these two critical points we
linearize the equations (\ref{dinamic1}) and (\ref{dinamic2}) near
these critical points. From the study of the eigenvalues of the
system we found that the first critical  point, Eq.~(\ref{crit1}),
could be a center or a saddle point, depending on the values of
the parameters of the model.
On the other hand, the second critical point Eq.~(\ref{crit2}) is
a saddle.

Stable static solutions correspond to a center. This imposes the
following conditions for the parameters ($B$ and $C$) in order
that the critical point, Eq.~(\ref{crit1}), becomes a center.

\begin{equation}B<0\,\,, -\frac{1}{64 B}< C < -\sqrt{\frac{3}{B^2}}-\frac{7}{4 B}.
\label{estac}
\end{equation}

These conditions also ensure that $x>0$ and the positivity of the
energy density. If we consider the definition of the parameters
$A$, $B$ and $C$ given in Eq.(\ref{C}) we can note that
Eq.~(\ref{estac}) is in agrement with the stability conditions
that were found previously.

\begin{figure}[h]
\begin{center}
\includegraphics[width=2.5in,angle=0,clip=true]{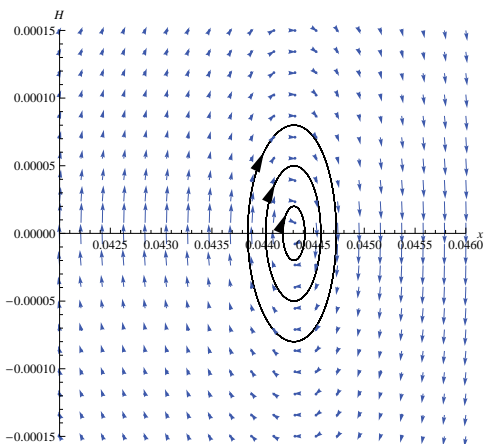}
\caption{Plot showing the Direction Field near the center critical
point and three numerical solution. Here we have used unit where
$\kappa = 1$.} \label{fase1}
\end{center}
\end{figure}

In Fig.~\ref{fase1} it is shown a phase portrait near the center
critical point ($H = 0, x = 0.0443$) for three numerical solution
to Eqs.~(\ref{dinamic1}) and (\ref{dinamic2}). Also, in this
figure we have included the \textit{Direction Field} of the system
in order to have a picture of how a general solution look like. In
this figure we have used the values $B = -1$ and $C = 0.016$.

Once a transition to a slow roll phase inflationary phase takes
place, we then have to see if the resulting inflationary phase can
provide enough e-foldings for the solution of the Big-Bang
standard problems.

We consider  the relevant equations  in the slow roll regime, i.e.
for $\dot{\phi}^2/2\ll V(\phi)$ and when the  scalar field $\phi$
is large, but finite. Dropping higher powers of $\dot{\phi}$ in
the contributions for the kinetic energy,  we obtain,
\begin{eqnarray}
\rho =  \frac{1}{2}\,\gamma(\phi)\,\dot{\phi}^2 + V_{eff}, \label{energydensityslow}\\
\end{eqnarray}
with
\begin{eqnarray}
\gamma(\phi)=1-\frac{2\delta
b_g(M^4e^{-2\alpha\phi/M_p}+V_1)}{4[b_g(M^4e^{-2\alpha\phi/M_p}+V_1)-V_2]},
\end{eqnarray}
and
\begin{eqnarray}
V_{eff}(\phi) =\frac{(M^4
e^{-2\alpha\phi/M_p}+V_1)^2}{4[b_gV_1-V_2]\left[1+\frac{b_gM^4e^{-2\alpha\phi/M_p}}{(b_gV_1-V_2)}\right]}.\label{gamma}
\end{eqnarray}
Here we have taken $s=1$, $b_g>0$ and $V_1<0$, then one  obtains
without fine tuning a vacuum state with zero energy density and
thus $V_{eff}(\dot{\phi}=0,\phi)\geq 0$.

In the slow roll approximation, we can drop the second derivative
term of $\phi$ and the second power of $\dot{\phi}$ in the
equation for $H^2$ and obtain,
\begin{eqnarray}
3H \gamma \dot{\phi} = -
V^{\prime}_{eff}\,,\,\,\,\mbox{and}\;\;\,\,\,\, 3H^2 = \kappa
V_{eff}, \label{Friedmannslow}
\end{eqnarray}
where $ V^{\prime}_{eff}= \frac{dV_{eff}}{d\phi}$. The relevant
expression for $V_{eff}$ will be that given by Eq.(\ref{gamma}),
i.e., where all higher derivatives are ignored, consistent with
the slow roll approximation.

We now display the relevant expressions for the asymptotic value
of $\phi$, these are
\begin{eqnarray}
V_{eff} \approx V_0+\overline{V}e^{-2\alpha\phi/M_p},
\end{eqnarray}
where
$$
V_0=\frac{V_1^2}{4(b_g\,V_1-V_2)}\,\,,\,\,\;\;\overline{V}=\frac{V_1\,M^4(b_gV_1-2V_2)}{4(b_g\,V_1-V_2)^2}\,,
$$
and
\begin{eqnarray}
 \gamma\approx\left[1-\frac{\delta
b_gV_1}{2(b_gV_1-V_2)}\right]+e^{-2\alpha\phi/M_p}\,\frac{\delta
b_gV_2M^4}{2(b_gV_1-V_2)^2}=\gamma_0+\gamma_1\,e^{-2\alpha\phi/M_p}.
\label{gamma0}
\end{eqnarray}
Note that $\overline{V}<0$ since $b_g>0$, $V_1<0$ and $V_2<0$.

At  the end of inflation, where, $\phi=\phi_{end}$, the parameter
$\varepsilon$, defined by $\varepsilon = - \frac{\dot{H}}{H^2}$,
takes an approximated  value equal to one  (analogous to
$\ddot{a}\simeq 0$ ).  The condition under which inflation takes
place can be summarized with the parameter $\varepsilon$
satisfying the inequality $\varepsilon < 1$ (or  $\ddot{a}>0$).
Taking the derivative with respect to the cosmic time of the
Hubble parameter and from Eq.(\ref{Friedmannslow}), we obtain that
the condition $\varepsilon \simeq1$ gives

\begin{equation}
 \varepsilon =\frac{1}{2\kappa\gamma}(V^{\prime}_{eff}/V_{eff})^2 \simeq 1, \label{endofinf.eq.}
\end{equation}
 working to leading order, setting $\gamma \approx\gamma_0$, $V_{eff} \approx V_0 $ and
 $V^{\prime}_{eff}\approx - (2\alpha/M_p)\, \overline{V} exp(-2\alpha \phi/M_p)$,
 we obtain
\begin{equation}
 e^{-2\alpha\phi_{end}/M_p}\simeq \frac{V_0\,M_p\sqrt{\kappa\,\gamma_0}}{\sqrt{2}\,\alpha\,\mid\overline{V}\mid}. \label{sol.endofinf.eq.}
\end{equation}
We now consider $\phi_{*}$ and the requirement that this precedes
$\phi_{end}$ by $N$ e-foldings,
\begin{equation}
N= \int_{t_*}^{t_{end}} Hdt \approx \int_{\phi_*}^{\phi_{end}}
\frac{H}{\dot{\phi}} d\phi \approx -\int_{\phi_*}^{\phi_{end}}
\frac{3H^2\gamma}{ V^{\prime}_{eff}} d\phi.
\end{equation}
In the following, the subscripts $*$ and $end$ are used to denote
to the epoch when the cosmological scale exit the horizon and the
end of inflation, respectively.

Solving $H^2$ in terms of $V_{eff}$ using
Eq.(\ref{Friedmannslow}), working to leading order, setting
$\gamma = \gamma_0$ and integrating, we obtain that the relation
between $\phi_{*}$ and $N$ becomes

\begin{equation}\label{sol.crossing.final}
e^{2\alpha \phi_{*}/M_p} \approx
\frac{\alpha\,\mid\overline{V}\mid}{\sqrt{\kappa\gamma_0}\,V_0\,M_p}\left[\sqrt{2}-\frac{4\alpha\,N}{\sqrt{\kappa\,\gamma_0}}\right].
\end{equation}

We finally calculate the power of the primordial scalar
perturbations.  The power spectrum of the curvature perturbation
in the slow-roll approximation for a not-canonically kinetic term
becomes Ref.\cite{Garriga}(see also Refs.\cite{p1})

\begin{equation}
P_S = k_1\,\frac{H^2}{c_s\,\varepsilon},\label{pec}
\end{equation}
where it was  defined the "speed of sound", $c_s$, as $
c_s^2=\frac{P_{,\,Z}}{P_{,\,Z}+2ZP_{,\,ZZ}}, $
 with $P(Z,\phi)$  function of the scalar field $\phi$ and  the
 kinetic term, $Z=(1/2)\widetilde{g}^{\,\mu\nu}\partial_\mu \phi\partial_\nu \phi$, and $k_1=(8\pi M_p^2)^{-1}$.
 Here $P_{,\,Z}$ denote the derivative with respect $Z$.
 In our case $P(Z,\phi)=\gamma(\phi)\,Z-V_{eff}$, with $Z=\dot{\phi}^2/2$.
 Thus, from Eq.(\ref{pec}) we get
\begin{equation}
P_S=k_1\frac{H^4}{\gamma(\phi) \dot{\phi}^2}.\label{eq10}
\end{equation}

 The scalar spectral index $n_s$ is defined by
\begin{equation}
n_s-1=\frac{d\ln P_S}{d\ln k}=-2\varepsilon-\eta-\xi,\label{ns}
\end{equation}
where $\eta=\frac{\dot{\varepsilon}}{\varepsilon\,H}$ and
$\xi=\frac{\dot{c_s}}{c_s\,H}$.

On the other hand, the generation of tensor perturbations during
inflation  would produce gravitational waves. The amplitude of the
tensor perturbations was evaluated in Ref.\cite{Garriga}. In our
case
\begin{equation}
P_T=\frac{2}{3\pi^2}\,\left(\frac{2ZP_{,\,Z}-P}{M_{P}^4}\right),\label{Pt}
\end{equation}
where the tensor spectral index, $n_T$, becomes $ n_T=\frac{d\ln
P_T}{d\ln k}=-2\varepsilon, $ and they satisfy a generalized
consistency relation given by $ r=\frac{P_T}{P_S}=-8\,c_s\,n_T. $

\begin{figure}[th]
\includegraphics[width=5.0in,angle=0,clip=true]{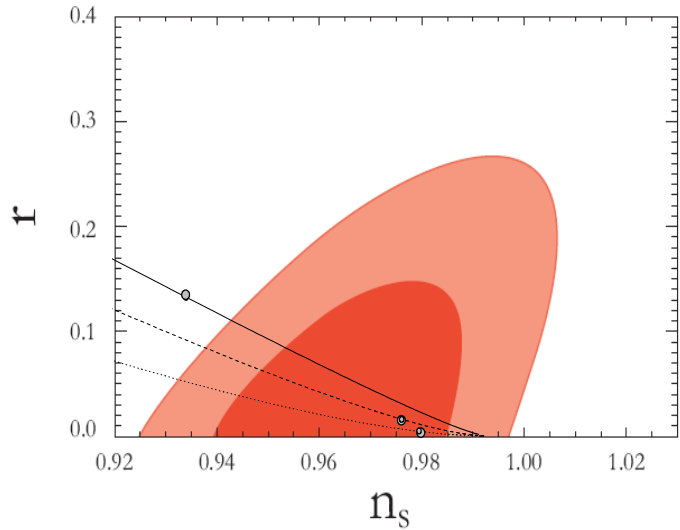}
\caption{ The plot shows $r$ versus $n_s$ for three values of
$\alpha$. For $\alpha=0.02$ solid line,  $\alpha=0.05$ dash line
and $\alpha=0.1$ dots line, respectively. Here, we have fixed the
values $y=0.52$, $M=1$ and $\kappa=1$, respectively. The  dots
represent the number of e-folds for the value N = 60. The
seven-year WMAP data places stronger limits on the tensor-scalar
ratio \cite{Larson:2010gs}.\label{fig2}}
\end{figure}

Therefore, the scalar field (to leading order) that should appear
in Eq. (\ref{eq10}) should be $\sqrt{\gamma_0}\phi$ and thus we
have
\begin{equation}\label{primordialpert.corr}
 P_S =k_1 \left(
 \frac{H^2}{\sqrt{\gamma_0}\dot{\phi}}\right)^2=
 \frac{\kappa^2\,V_0\,k_1}{12}\,\left[\sqrt{2}-\frac{4\alpha\,N}{M_p\sqrt{\kappa\,\gamma_0}}\right]^2.
 \end{equation}
This quantity should be evaluated at $\phi=\phi_{*}$ given by
Eq.(\ref{sol.crossing.final}).

In Fig.\ref{fig2} we show the dependence of the tensor-scalar
ratio $r$ on the spectral index $n_s$. From left to right
$\alpha=0.01$ (solid line), $\alpha=0.05$ (dash line) and
$\alpha=0.1$ (dots line), respectively. The   dots  represent the
number of e-folds for the value N = 60. From
Ref.\cite{Larson:2010gs}, two-dimensional marginalized
 constraints (68$\%$ and 95$\%$ confidence levels) on inflationary parameters
$r$ and $n_s$, the spectral index of fluctuations, defined at
$k_0$ = 0.002 Mpc$^{-1}$. The seven-year WMAP data places stronger
limits on $r$. In order to write down values that relate $n_s$ and
$r$, we used Eqs.(\ref{ns}), (\ref{Pt}) and
(\ref{primordialpert.corr}).  Also we have used the values
$y=0.52$, $M=1$ and $\kappa=1$, respectively. We noted that the
parameter $\alpha$, which lies in the range $1 > \alpha > 0$, the
model is well supported by the data as could be seen from
Fig.\ref{fig2}.

The dilaton $\phi$ dependence of the effective Lagrangian appears
only as a result of the spontaneous breakdown of the scale
invariance. If no fine tuning is made, the energy density
$\rho(\phi,Z)$ and the pressure $p(\phi,Z)$ depend quadratically
upon the kinetic term $Z$. Hence TMT represents an explicit
example of {\it the effective} $k$-{\it essence resulting from
first principles without any exotic term} in the underlying action
intended for obtaining this result. These non linearities in
$\rho(\phi,Z)$ and $p(\phi,Z)$ play a crucial role in existence
and stability of the emerging universe solutions. In this paper we
have been successful in describing an emergent universe in a TMT
sort of theory.

\section{Acknowledgements}
This work was supported by Comision Nacional de Ciencias y
Tecnolog\'{\i}a through FONDECYT  Grants 1110230 (SdC), 1090613
(RH). Also it was supported by Pontificia Universidad Catolica de
Valparaiso  through grants 123.787-2007 (SdC) and 123.703-2009
(RH). One of us (E.I.G) would like to thank the astrophysics and
cosmology group at the Pontificia Universidad Catolica de
Valparaiso for hospitality and the Grant FSM0806 Programa MECE 2
Educacion Superior Fondo de Innovacion Visitas de Especialistas.
P. L. is supported by FONDECYT grant N$^0$ 11090410 and by
Direcci\'on de Investigaci\'on de la Universidad del
B\'{\i}o-B\'{\i}o through Grant N$^0$ 096207 1/R.

\end{document}